# فناوری‌های تجدیدپذیر: مزایا، معایب و سیاست‌های راهبردی

حمید ذاکرنژاد- علی سلیمانی- میلاد محبتی

دانشگاه شهید بهشتی تهران

[al.soleimani@mail.sbu.ac.ir](mailto:al.soleimani@mail.sbu.ac.ir) – [h.zakernejad@mail.sbu.ac.ir](mailto:h.zakernejad@mail.sbu.ac.ir)

## چکیده

رشد روز افزون نیاز به انرژی و پایان‌پذیری سوخت‌های فسیلی و همچنین افزایش آلودگی‌های زیست محیطی، سبب شده تا محققان در جستجوی جایگزینی برای این منابع در چرخه‌ی انرژی باشند. در این میان منابع انرژی تجدیدپذیر و پاک حائز اهمیت می‌باشند. به‌طوری‌که برای تأمین تقاضای انرژی و همچنین جهت رشد و توسعه اقتصادی، به‌ویژه در کشورهای درحال توسعه، پتانسیل بسیار بالایی دارند. در این مقاله به بررسی انواع فناوری‌های مرسوم انرژی‌های تجدیدپذیر به تفکیک پرداخته شده است. تغییر از فناوری فسیلی به تولید مبتنی بر انرژی‌های دوست‌دار طبیعت، زمان بر است و نیاز به هزینه سرمایه‌گذاری بسیار بالایی دارد. همچنین تولید انرژی‌های نو ممکن است باعث غیر فعال شدن برخی نیروگاه‌های فسیلی گردد که در نتیجه بیکاری و رشد منفی اقتصادی به همراه خواهد داشت. در ادامه‌ی این پژوهش، خط مشی کشورهای مختلف مورد مطالعه قرار گرفته و راهکارهایی جهت گسترش این منابع پیشنهاد شده است.

## کلید واژه

آلاینده‌های زیست محیطی- انرژی‌های تجدیدپذیر- بازار انرژی- چشم‌انداز ۲۰ ساله- سیاست دولت‌ها- شبکه‌ی هوشمند

## مقدمه

امروزه شهرهای مختلف جهان به مصرف‌کنندگان بی‌رویه منابع طبیعی و تولیدکنندگان بزرگ زباله و ضایعات تبدیل شده‌اند. ایران نیز از این قاعده مستثنی نیست. برای مثال کلان‌شهری همچون تهران با اختصاص تنها ۱۲٪ از مساحت کشور، ۱۵٪ از حجم آلودگی منتشره در هوا را در سطح کشور به خود اختصاص داده است [۱]. افزایش رشد جمعیت و نیاز روز افزون بشر به انرژی از یک طرف و پایان‌پذیری سوخت‌های فسیلی و مشکلات زیست محیطی، از طرفی دیگر سبب گرایش برنامه‌ریزان و سیاست‌گذاران کشور به توسعه منابع قابل جایگزینی با سوخت‌های فسیلی شده است. در طی سال‌های اخیر تحقیقات بسیاری در زمینه‌ی استفاده هر چه بیشتر از منابع پایدار و تجدیدپذیر صورت گرفته است. منظور از انرژی‌های تجدیدپذیر، انرژی‌هایی هستند که به طور مستقیم توسط طبیعت و نیروی خورشید، (مانند حرارت، فتوالکتریک و فتوشیمی) یا به صورت غیرمستقیم (مانند باد، برق‌-آبی و زیست توده) تولید می‌شوند [۲]. با توجه به برخورداری از پتانسیل مطلوب و مناسب انرژی‌های تجدیدپذیر در کشور، توسعه این منابع ارزشمند و خدادادی کاملاً ضروری به نظر می‌رسد؛ چراکه از این طریق می‌توان در جهت اهداف زیست محیطی نیز گام برداشت. جدول ۱ رشد تولید منابع تجدیدپذیر بین سال‌های ۲۰۱۰ تا ۲۰۳۵ را نشان می‌دهد. همچنین در شکل ۱، سهم این منابع در بخش‌های مختلف برای سال‌های فوق پیش‌بینی شده است [۳].

در کنار مزایای فراوان، استفاده از این منابع با چالش‌هایی همراه است. مشکل انرژی‌هایی همانند باد و خورشید، عدم دسترسی مداوم و غیرقابل پیش‌بینی بودن آن‌هاست. علی‌رغم نیروگاه‌های سنتی، این منابع انرژی قابل برنامه‌ریزی نیستند و توان خروجی آنها غیر قابل کنترل است [۴، ۵]. نکته‌ی دیگر اینکه استفاده بیش از حد از این منابع در تأمین بار شبکه، موجب بروز مشکلاتی از قبیل عدم تثبیت ولتاژ، مشخص نبودن میزان دقیق تولیدات و افزایش نااطمینانی سیستم می‌گردد [۶]. همچنین تغییر از فناوری فسیلی به

فناوری تولید انرژی‌های تجدیدپذیر زمان بر است و نیاز به هزینه سرمایه‌گذاری بسیار بالایی دارد. در این راه، اعطای مشوق‌های مالی، ایجاد صندوق‌های حمایت مالی از این منابع توسط دولت و ایجاد بستر و شرایط مناسب جهت توسعه این صنعت در کشور می‌تواند راهگشا باشد. در کشورهایی که دولت بازیگر اصلی در بخش انرژی است، برنامه‌ها و راهبردهای ملی با افق‌های مشخص برای پیشرفت منابع تجدید پذیر صورت می‌گیرد [۷, ۸].

جدول ۱: تولید جهانی منابع تجدیدپذیر بر اساس نوع آن‌ها

| | ۲۰۱۰ | ۲۰۲۰ | ۲۰۳۵ |
|---|---|---|---|
| مجموع تولید برق (Twh) | ۴۲۰۶ | ۶۹۹۹ | ۱۱۳۴۲ |
| انرژی زیستی | ۳۳۱ | ۶۹۶ | ۱۴۸۷ |
| انرژی برق آبی | ۳۴۳۱ | ۴۵۱۳ | ۵۶۰۷ |
| انرژی بادی | ۳۴۲ | ۱۲۷۲ | ۲۶۸۱ |
| زمین گرمایی | ۶۸ | ۱۳۱ | ۳۱۵ |
| سلول خورشیدی | ۳۲ | ۳۳۲ | ۸۴۶ |
| نیروگاه متمرکز خورشیدی | ۲ | ۵۰ | ۲۷۸ |
| سهم تجدید پذیر از کل انرژی | ۳۱٪ | ۲۵٪ | ۲۰٪ |

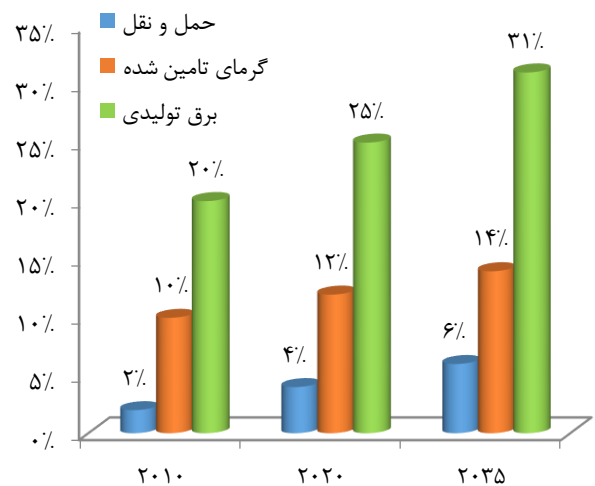

شکل ۱: سهم منابع تجدیدپذیر در بخش‌های مختلف

با توجه به نکات اشاره شده، این مقاله در بخش ۱ به معرفی انواع منابع تجدیدپذیر می‌پردازد. در بخش ۲ مزایا و چالش‌های استفاده از فناوری‌های مذکور مورد بررسی قرار گرفته است. سیاست‌گذاری کشورهای مختلف و چشم‌انداز آن‌ها در توسعه‌ی این منابع در بخش ۳ بیان شده است. بخش ۴ نیز به نتیجه‌گیری و بحث نهایی می‌پردازد.

## ۱- معرفی انواع منابع تجدیدپذیر

منابع تجدیدپذیر فراوانی در جهان موجود است. مزیت اصلی این منابع انرژی، رایگان و در دسترس بودن آن‌هاست. در این بخش به معرفی فناوری‌های مرسوم و ظرفیت تولید هرکدام در سراسر جهان می‌پردازیم.

### ۱-۱- انرژی خورشیدی

تولید انرژی خورشیدی شامل استفاده از انرژی خورشید برای گرم کردن آب با سیستم حرارتی خورشید و یا تولید الکتریسیته توسط سلول فتوولتائیک[1] و سیستم متمرکزکننده توان خورشیدی[2] می‌باشد [۹]. کارایی این فناوری‌ها در چند دهه گذشته با نصب تعداد زیادی از آن‌ها در اقصی نقاط جهان اثبات شده است [۱۰].

### ۱-۱-۱- سیستم فتوولتائیک

سیستم فتوولتاییک، انرژی خورشید را بدون رابط گرمایی، مستقیماً به الکتریسیته تبدیل می‌نماید. ساختمان اصلی این سیستم شامل سلول فتوولتائیک می باشد. تمامی سلول‌های خورشیدی به ماده جذب کننده نور در ساختار خود نیاز دارند که فوتون‌ها را جذب کرده و با خاصیت فتوولتائیک، الکترون های آزاد تولید نمایند[۱۱].

سلول‌های خورشیدی در یک ماژول با توان ۵۰ تا ۲۰۰ وات کنار هم قرار می‌گیرند. ماژول‌های PV با مجموعه ای از ابزارهای الکترونیک قدرت (اینورترها ، باتری‌ها و دیگر تجهیزات الکتریکی) به‌هم مرتبط می‌گردند. بزرگترین مزیت این سیستم‌ها، ماژولار بودن آن‌هاست. بدان معنا که ماژول‌ها

---

[2] – Concentrated solar Power          [1] – PhotoVoltaic Cell

می‌توانند در کنار یکدیگر قرار گیرند و از چند وات تا ده‌ها مگاوات توان تولید نمایند [2].

مرسوم‌ترین فناوری سلول‌های خورشیدی، سیستم‌های مبتنی بر سیلیکون می‌باشند که به دو فناوری مونوکریستال[1] و پلی‌کریستال[2] تقسیم می‌شوند. فناوری جدید این سلول‌ها فیلم نازک[3] نامیده می‌شود که از نیمه‌هادی‌های غیر سیلیکونی تشکیل می‌شوند و به دلیل ویژگی‌های منحصر به فرد خود، از اهمیت بالایی در این حیطه برخوردارند. نوع مونو کریستال، قدیمی‌ترین فناوری سلول‌های خورشیدی است و تا این زمان پیشرفت بسیاری داشته است. همچنین در بین فناوری‌های موجود بیشترین بازده را دارد. جدیدترین آن‌ها حدود 22 درصد از انرژی تابیده شده را به الکتریسیته تبدیل می‌کنند. عمر این سلول‌ها در مقایسه با دیگر فناوری‌ها بالاتر است اما گران‌ترین در بین این فناوری‌ها هستند. پنل‌های پلی‌کریستال به دلیل خلوص کمتر، بازدهی در حدود 13 تا 16 درصد دارند و هزینه تمام شده آن‌ها پایین‌تر می‌باشد. فناوری فیلم نازک ارزان‌تر از دو مورد معرفی شده می‌باشد. اما بازده آن‌ها تنها 9 درصد است. با این وجود ویژگی ضخامت بسیار کم آن‌ها باعث شده تا برای موارد خاص قابل استفاده باشند. همچنین می‌توان آن‌ها را به شکل انعطاف‌پذیر ساخت [12].

سیستم‌های فتوولتائیک به دو شکل جزیره‌ای و متصل به شبکه قابل بهره‌برداری هستند. سیستم‌های متصل به شبکه، اقتصادی‌ترند زیرا نیازی به نصب باتری ندارند و توان خود را مستقیماً به شبکه تزریق می‌کنند. توان نوع مسکونی بین 1 تا 4 کیلووات و برای نوع سقفی و تجاری از 10 کیلووات تا چندین مگاوات می‌باشد [13].

تاکنون از سلول‌های خورشیدی در تولید انرژی، پمپاژ آب، ارتباطات، ماهواره‌ها و ربات‌های فضاپیما استفاده شده است. با چنین کاربردی سالانه تقاضا برای سلول‌های خورشیدی رو به افزایش است. در شکل 2 سهم هر کشور از تولید سیستم‌های فتوولتائیک نصب شده در جهان تا پایان سال 2016

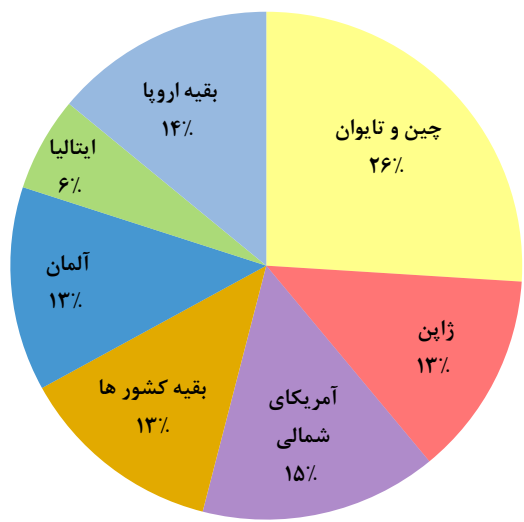

شکل 2: سهم کشورهای مختلف در تولید سلول‌های خورشیدی

نشان داده شده است. با توجه به شکل، عمده تولید سلول و مونتاژ ماژول، در چین و تایوان صورت می‌گیرد [14].

### 1-1-2- سیستم متمرکزکننده توان خورشیدی

فناوری متمرکزکننده توان خورشیدی، با متمرکز کردن نور خورشید سبب گرم شدن یک سیال می‌گردد. این سیال سبب چرخاندن توربین و تولید انرژی برق می‌شود. بازه تولید توان این منابع می‌تواند از چند ده کیلووات تا صدها مگاوات باشد [15, 16].

### 1-1-3- سیستم‌های تهویه مطبوع خورشیدی[4]

این فناوری انرژی گرمایی خورشید را جمع‌آوری کرده و از آن برای تولید آب گرم، گرمایش و سرمایش مناطق مسکونی استفاده می‌کند. کشورهایی که بیشترین بهره را از این فناوری برده‌اند شامل چین، آلمان، ترکیه، برزیل و هند می‌باشند [17].

### 1-2- انرژی باد

انرژی باد به دلیل فراوانی، تجدیدپذیری، پاک بودن و اشغال فضای کم جایگزینی مناسب برای انرژی فسیلی می‌باشد [17]. انرژی باد با استفاده از توربین‌های بادی قابل تبدیل به انرژی الکتریکی است. اولین توربین بادی برای استفاده از انرژی برق در اوایل قرن بیستم اختراع شد. از دهه

---

[1] – Mono-Crystalline
[2] – Poly-Crystalline
[3] – Thin Film
[4] – Solar thermal heating and cooling

۱۹۷۰ تا ۱۹۹۰ بیشرفت‌های بسیاری حاصل شد و امروزه توربین‌های بادی به یکی از مهم‌ترین فناوری‌های پایدار و تجدیدپذیر بدل شده‌اند [۱۸]؛ به طوری‌که در سال ۲۰۱۱ بیش از ۲۵٪ انرژی دانمارک از این راه تامین گردید. امروزه با پیشرفت فناوری توربین‌های بادی، هزینه به ازای هر کیلووات ساعت نیروگاه بادی، حدوداً با نیروگاه گازی یا زغال سنگی برابری می‌کند [۵].

توربین‌های بادی در خشکی در کنار یک دیگر نیروگاه‌های بادی یا مزارع بادی را تشکیل می‌دهند. این مزارع بادی معمولاً توانی بین ۵ تا ۳۰۰ مگاوات تولید می‌کنند. فناوری توربین های بادی شامل دو نوع دریایی[1] و ساحلی[2] می‌باشد. نوع دریایی این فناوری پیشرفت کمتری داشته و دارای هزینه سرمایه‌گذاری و تعمیرات بالاتری است [۱۹, ۲۰].

در طی پنج سال اخیر، بازار جهانی توربین‌های بادی رشدی بالغ بر ۳۰٪ در سال داشته است و توان بادی نقش فزاینده‌ای در تولید برق، خصوصاً در کشورهایی مانند آلمان و اسپانیا ایفا می‌کند [۲۱].

### ۱-۲-۱- فناوری توربین بادی

توربین های بادی به دوسته اصلی تقسیم می شوند: توربین های سرعت ثابت[3] و توربین های سرعت متغیر[4] [۲۲]. توربین‌های سرعت ثابت توسط یک جعبه دنده چند مرحله‌ای و یک ژنراتور سنکرون، انرژی الکتریکی تولید می‌کنند. در این فناوری ژنراتور مستقیماً توسط ترانسفورماتور به شبکه برق متصل می‌گردد. قیمت نسبتاً پایین، سادگی و قدرتمندی این نوع توربین‌ها سبب تجاری شدن آن‌ها شده است. این پیکربندی شامل ایراداتی از جمله کنترل کیفیت توان پایین، عدم کنترل توان راکتیو و عدم وجود کنترل سرعت برای رسیدن به کارایی بهینه آیرودینامیکی، می‌باشد [۲۳].

از طرف دیگر استفاده از توربینی که بتواند در سرعت‌های مختلف تولید داشته باشد، می‌تواند بازدهی را افزایش دهد. در توربین‌های سرعت متغیر، خروجی کاملاً به آیرودینامیک توربین و وضعیت باد بستگی دارد. فناوری اخیر سالانه ۱۰٪ انرژی بیشتری نسبت به توربین سرعت ثابت تولید می‌نماید.

این توربین‌ها به ادوات الکترونیک قدرت پیشرفته نیازمندند که همین امر موجب افزایش قیمت کلی آن‌ها شده است. مشکل دیگر این پیکربندی، وجود هارمونیک‌های بالا در آن می‌باشد [۲۴].

### ۱-۳- انرژی برق آبی

انرژی برق آبی انرژی‌ای است که از حرکت آب بدست می‌آید و به وسیله توربین به انرژی برق تبدیل می‌شود. نیروگاه‌های برق آبی برای تبدیل انرژی مکانیکی به الکتریکی، به سه دسته تقسیم می‌شوند [۲۵]:

۱) سدهای مرتفع
۲) نیروگاه تلمبه ذخیره‌ای
۳) جریان رودخانه‌ای

در این میان متداول‌ترین راه، استفاده از سد می‌باشد. نیروگاه‌های برق آبی رنج تولیدی گسترده‌ای ازچند وات تا چند گیگاوات را شامل می‌شوند. عظیم‌ترین سدهای احداث شده در دنیا، Itaipu Dam در برزیل با توان ۱۴ گیگاوات و سد Three Gorges Dam یا سه گلوی چین با توان تولیدی ۲۲٫۴ گیگاوات می‌باشند. هر دوی این سدها، سالانه بین ۸۰ تا ۱۰۰ تراوات ساعت انرژی به شبکه تزریق می‌نمایند [۲۶-۲۸]. پنج کشور چین، برزیل، کانادا، ایالات متحده، و روسیه نیمی از انرژی برق آبی دنیا را تولید می‌کنند [۲۹]. در آینده شبکه‌های هوشمند، این نیروگاه‌ها می‌توانند نقش مهمی در ذخیره انرژی ایفا کنند.

### ۱-۳-۱- نیروگاه تلمبه ذخیره‌ای

نیروگاه تلمبه ذخیره‌ای نوعی نیروگاه است که برای استفاده از برق مازاد بر مصرف و بازتولید این برق در زمان‌هایی که تقاضای مصرف زیاد است به کار می‌رود. نیروگاه تلمبه ذخیره‌ای امکان افزایش ظرفیت تولید را از طریق تلمبه‌ی آب از یک مخزن پایین‌تر به یک مخزن بالاتر در مواقعی که تقاضای مصرف برق کم است (مثلا هنگام شب)، می‌دهد. این آب بعدها می‌تواند در مواقع مورد نیاز به مخزن پایین برگردد تا توربین‌ها را بچرخاند و در نهایت ژنراتور را به گردش درآورد [۳۰]. نمونه‌ی این نیروگاه‌ها در ایران، نیروگاه سیاه بیشه با

---

[1] off shore
[2] on shore
[3] fixed speed wind turbine
[4] Variable speed wind turbine

ظرفیت تولیدی ۱۰۴۰ مگاوات و ظرفیت پمپاژ ۹۴۰ مگاوات می‌باشد [۳۱].

### ۱-۳-۲- نیروگاه جریان رودخانه‌ای

نیروگاه‌های جریان رودخانه‌ای نوعی نیروگاه برق آبی هستند که فاقد ذخیره‌سازی آب یا ذخیره‌سازی حجم محدودی می‌باشند؛ که در حالت دوم به این مخزن حوضچه نیز می‌گویند. از مزایای این نوع نیروگاه‌ها، تولید انرژی تجدیدپذیر با حداقل تاثیرات محیطی و تغییرات اقلیم مناطق پایین دست می‌باشد. از نمونه‌های بزرگ ساخته شده می‌توان به نیروگاه Belo Monte Dam در برزیل با ظرفیت ۱۱٫۲ گیگاوات و Chief Joseph در آمریکا با ظرفیت ۲٫۶ گیگاوات اشاره کرد.

### ۱-۴- انرژی زمین گرمایی[۱]

انرژی زمین گرمایی شیوه‌ای کارآمد برای دریافت انرژی تجدیدپذیر از زمین، طی یک فرآیند طبیعی می‌باشد. این روش می‌تواند در ابعاد کوچک برای تامین گرمای منازل مسکونی یا یک پمپ گرمایی یا در ابعاد عظیم برای تولید انرژی در نیروگاه برق بکار رود. انرژی زمین گرمایی، یک منبع انرژی با هزینه مناسب، مطمئن و دوستدار محیط زیست است [۳۲]. انرژی زمین گرمایی از گرمای درونی زمین در اعماق سنگ‌ها که در گاز و آب محبوس شده است، حاصل می‌شود. دمای پوسته‌های زیرین زمین در موقعیت‌های جغرافیایی مختلف، متفاوت است. بسیاری از این سیستم‌ها با دمای بالا (بالاتر از ۱۸۰ درجه سانتی‌گراد) در نزدیکی آتش‌فشان‌ها واقع شده‌اند. این سیستم در مناطق با دمای میانی (بین ۱۰۰ تا ۱۸۰ درجه سانتی گراد) و دمای پایین (کمتر از ۱۰۰ درجه) نیز در ابعاد کوچکتر قابل استفاده هستند [۳۳, ۳۴]. از کابردهای این انرژی می‌توان به بهبود سیستم حمل و نقل الکتریکی و برفروبی و قابل تردد کردن مسیرهای صعب العبور اشاره کرد.

### ۱-۵- انرژی زیست‌توده[۲]

انرژی زیست‌توده شامل سوخت‌های مایع یا گازی (از قبیل اتانول، بیوتانول، بیودیزل) و سوخت‌های جامد (نظیر توده ذرت) است که جهت تولید الکتریسیته مورد استفاده قرار می‌گیرند. انرژی زیست‌توده تجدیدپذیر و پایدار می‌باشد. سوخت‌های جامد را می‌توان طی فرآیندی به سوخت‌های مایع یا گازی (سوخت زیستی) تبدیل کرد. سوخت زیستی[۳] بعد از تبدیل، قابل جابجایی و ذخیره‌سازی است که جهت پشتیبانی از انرژی‌های متناوب مانند باد بسیار کاربردی می‌باشد. بر این اساس انتظار می‌رود که زیست‌توده نقش مهمی در آینده تامین انرژی ایفا نماید [۳۵]. از جمله موارد استفاده از این انرژی، تولید برق، اتومبیل‌های بیوگاز سوز، روشنایی معابر و سوخت خودروها است.

### ۱-۶- انرژی هیدروژن

هیدروژن عمده‌ترین گزینه مطرح به عنوان حامل جدید انرژی و جایگزینی مناسب برای سوخت‌های فسیلی است. وسیله‌ای که با استفاده از آن می‌توان از هیدروژن انرژی گرفت، پیل سوختی[۴] نام دارد. پیل سوختی انرژی شیمیایی سوخت را مستقیماً به انرژی الکتریکی تبدیل می‌کند. برخلاف باتری‌ها که به علت محدود بودن مقدار ماده‌ی واکنش دهنده در مخزن باتری، پس از مدتی نمی‌توانند انرژی لازم را تأمین کنند، در پیل سوختی مواد واکنش دهنده به صورت پیوسته وارد پیل شده و فرآورده‌ها به صورت پیوسته خارج می‌شوند، بنابراین پیل سوختی می‌تواند به صورت پیوسته عمل نماید. همچنین به دلیل اینکه تبدیل انرژی به طور مستقیم روی می‌دهد؛ از بازدهی بالایی برخوردار است. در پیل سوختی گاز هیدروژن به عنوان سوخت مصرف شده و از واکنش آن با اکسیژن، علاوه بر انرژی الکتریکی، آب و حرارت نیز تولید می‌گردد. به عبارت دیگر در این تبدیل، عکس واکنش الکترولیز آب رخ می‌دهد. هیدروژن بهترین و ساده‌ترین سوخت جهت جایگزینی خودروهای احتراقی بوده و دارای راندمان به مراتب بالاتری نسبت به آن‌ها می‌باشد. پسماند حاصل از احتراق هیدروژن، بخار آب است که

---

[۱] – Geothermal Energy
[۲] – Biomass Energy
[۳] – Biofuel
[۴] – Fuel cell

مشکلات آلودگی در پی ندارد. همین امر تأثیرات به سزایی در کاهش آلودگی زیست محیطی شهرها ایفا می‌نماید.

از جمله معایب این فناوری می‌توان به گران بودن پیل‌های سوختی اشاره کرد که تولید انبوه آن را با مشکل مواجه می‌سازد. همچنین ممکن است در مدت طولانی کار، گرما موجب بروز مشکلاتی چون ناسازگاری عناصر و افت انرژی گردد [۳۶, ۳۷]. طبقه بندی رایج پیل‌های سوختی بر اساس الکترولیت آن‌ها می‌باشد [۳۸]:

۱. پیل سوختی پلیمری (PEMFC)
۲. پیل سوختی قلیایی (AFC)
۳. پیل سوختی اسید فسفریک (PAFC)
۴. پیل سوختی کربنات مذاب (MCFC)
۵. پیل سوختی اکسید جامد (SOFC)
۶. پیل سوختی متانولی (DMFC)

شکل ۳ سهم منابع تجدیدپذیر از کل تولیدات شبکه را برای کشورهای منتخب نمایش می‌دهد [۳۹].

## ۱-۷- سیستم‌های انرژی تجدیدپذیر هیبریدی[1]

سیستم‌های انرژی تجدیدپذیر هیبریدی، ترکیبی از منابع انرژی تجدیدپذیر و سنتی یا ترکیبی از چند منبع انرژی تجدیدپذیر می‌باشند که در حالت جزیره‌ای یا متصل به شبکه کار می‌کنند. سیستم‌های هیبریدی می‌توانند بر مشکل غیرقابل پیش‌بینی بودن منابع تجدیدپذیر، غلبه کنند. این سیستم‌ها دارای بازدهی بالاتری نسبت به حالتی که تنها از یک منبع انرژی استفاده شود، می‌باشند. شکل ۴ اجزای یک سیستم هیبریدی را نمایش می‌دهد.

## ۲- منابع تجدیدپذیر: مزایا و چالش‌ها

همانطور که در بخش‌های قبل اشاره شد، در کنار مزایای فراوان بهره‌مندی از منابع تجدیدپذیر، با برخی چالش‌ها جهت توسعه‌ی آن‌ها در شبکه مواجه هستیم. در این بخش به معرفی برخی مزایا و مشکلات پیشرو جهت استفاده از این منابع می‌پردازیم.

## ۲-۱ کاهش آلاینده‌های زیست محیطی

رشد روز افزون تقاضای انرژی و افزایش آلودگی‌های زیست محیطی سبب شده تا محققان و سرمایه‌گذاران، به سمت تأمین انرژی از منابع تجدیدپذیر ترغیب شوند. تولید انرژی از این منابع در مقایسه با معادل فسیلی آن‌ها، می‌تواند به طور قابل توجهی باعث کاهش انتشار گازهای گلخانه‌ای گردد [۴۰]. شکل ۵ نشان دهنده‌ی میزان انتشار $CO_2$ ناشی از تولید یک کیلووات ساعت توسط دو تکنولوژی پیشتاز یعنی، فتوولتائیک و گاز طبیعی است. اطلاعات استفاده شده از مرکز

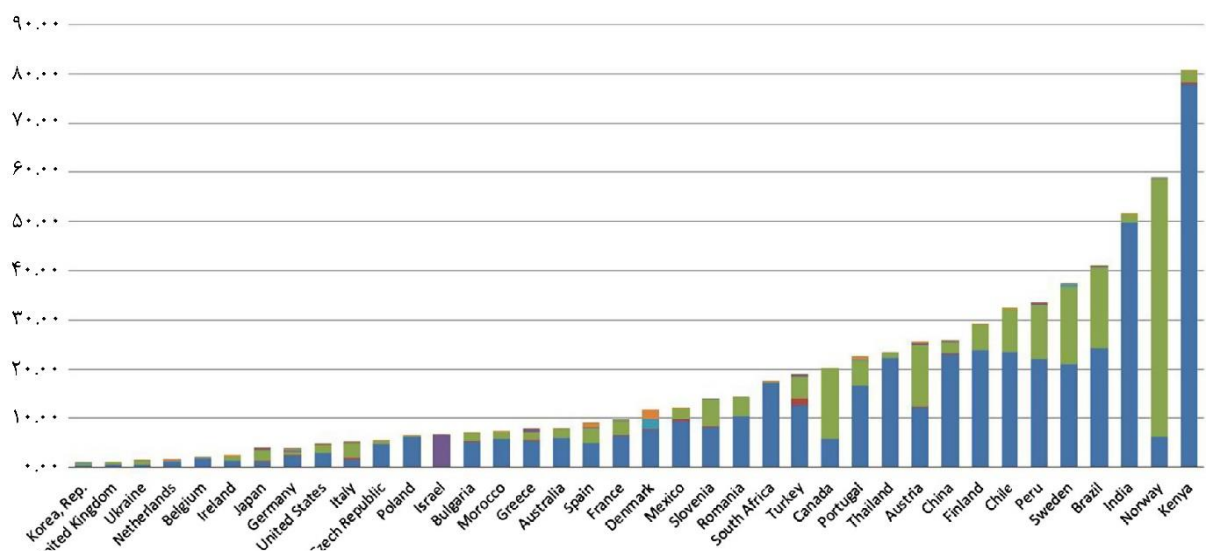

شکل ۳: سهم منابع تجدیدپذیر از کل تولیدات شبکه در کشورهای مختلف

---
[1] Hybrid Renewable Energy Systems

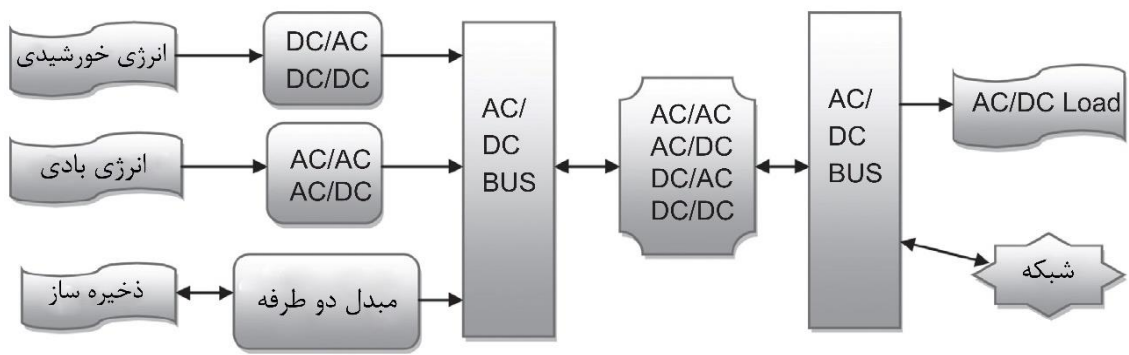

شکل ۴: اجزای اصلی سیستم هیبریدی تجدید پذیر خورشیدی- بادی

اطلاعات انرژی باز[1]گردآوری و در مقالات متعددی منتشر شده است [۴۱-۴۳].

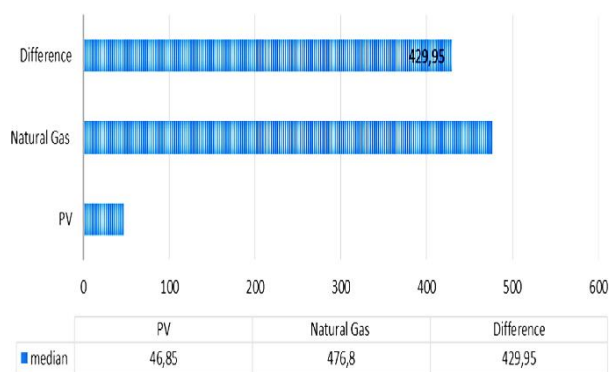

شکل ۵: مقایسه‌ی تولید گاز $CO_2$ توسط نیروگاه گازی و فتوولتائیک

همان طور که در شکل ۵ ملاحظه می‌شود، تفاوت میزان انتشار $CO_2$ برای تولید یک کیلووات ساعت انرژی بین دو فناوری فتوولتائیک و گاز طبیعی ۴۲۹٫۹۵ گرم خواهد بود. با توجه به فصل ۹ گزارش ویژه IPCC درباره منابع انرژی تجدیدپذیر و مقابله با تغییرات اقلیمی، مقادیر متوسط تولید گازهای گلخانه‌ای برای فن آوری‌های تجدیدپذیر بین ۴ تا ۴۶ گرم گاز $CO_2$ به ازای هر کیلووات ساعت می‌باشد. در حالی که این مقدار برای واحدهایی که از سوخت‌های فسیلی استفاده می‌کنند، بین ۴۵۰ تا ۱۰۰۰ گرم در هر کیلووات ساعت است [۴۰].

طبق برآورد آژانس بین‌المللی انرژی[2]، سهم انرژی‌های تجدیدپذیر از تولید برق در خوش‌بینانه‌ترین سناریو، از ۱۸٫۳٪ در سال ۲۰۰۲ به ۳۹٪ تا سال ۲۰۵۰ افزایش خواهد یافت. به منظور کاهش انتشار $CO_2$ در جهان به میزان ۵۰٪ تا سال ۲۰۵۰، انرژی‌های تجدیدپذیر در محدود کردن رشد میانگین دمای جهانی بین ۲ تا ۲٫۴ درجه سانتی‌گراد، نقشی کلیدی را ایفا خواهند نمود [۴۴].

## ۲-۲- برق‌رسانی به مناطق دور افتاده

در حال حاضر بیش از ۲ میلیارد از جمعیت زمین در روستاها و جزایز مسکونی دور افتاده ساکن هستند که به شبکه‌ی سراسری برق دسترسی ندارند [۴۵]. راه حل مطلوب در چنین مناطقی جهت برق‌رسانی و همچنین جلوگیری از انتشار آلاینده‌های سوخت‌های فسیلی، استفاده از انرژی‌های تجدیدپذیر موجود در منطقه است [۴۶]. بسیاری از متخصصان بر این باورند که فناوری‌هایی همچون باد، خورشید و نیروگاه آبی کوچک نه تنها اقتصادی، بلکه برای مناطق دورافتاده ایده آل هستند [۴۷].

## ۲-۳- تأثیر توسعه‌ی انرژی‌های تجدیدپذیر در رفاه عمومی

براساس مطالعه‌ی انجمن بین‌المللی انرژی‌های تجدیدپذیر در سال ۲۰۱۶، تأثیر توسعه‌ی انرژی‌های تجدیدپذیر بر رفاه جهانی مثبت بوده و شاهد افزایش رفاه جهانی به میزان ۲٫۷٪ در سال ۲۰۳۰ خواهیم بود. بر اساس برآورد انجمن بین‌المللی انرژی‌های تجدیدپذیر، بیشترین میزان بهبود رفاه ناشی از توسعه‌ی منابع تجدیدپذیر در سال ۲۰۳۰ مربوط به کشورهای هند، اوکراین، آمریکا، استرالیا، اندونزی، آفریقای جنوبی، چین و ژاپن خواهد بود [۴۸, ۴۹].

---

[1] – Open Energy Information
[2] - International Energy Agency

## 2-4- تأثیر توسعه‌ی انرژی‌های تجدیدپذیر بر اشتغال

نتایج بررسی‌ها بیانگر آن است که توسعه‌ی انرژی‌های تجدیدپذیر سبب ایجاد بیش از 8,1 میلیون شغل به صورت مستقیم و غیرمستقیم در سال 2015 در جهان شده که سهم کشورهای آسیایی بیش از 60٪ بوده است. علاوه بر این، با احداث نیروگاه‌های برق آبی بزرگ، بیش از 1,3 میلیون شغل مستقیم در سال 2015 به وجود آمده است [48]. استمرار سیاست‌های اتخاذ شده از سوی کشورهای جهان در زمینه‌ی توسعه‌ی انرژی‌های تجدیدپذیر، تضمین کننده‌ی افزایش تعداد شاغلان در این بخش خواهد بود. به عنوان مثال، هند اقدام به برنامه‌ریزی جهت تولید برق خورشیدی به میزان 100 هزار مگاوات و اشتغال 515 هزار نفر در سال 2022 کرده است [50].

## 2-5- چالش‌های توسعه‌ی منابع تجدیدپذیر

در کنار مزایای فراوان، جهت اتصال منابع تجدیدپذیر به شبکه با مشکلاتی از جمله یکپارچه‌سازی این منابع مواجه هستیم، چرا که شبکه‌ها به طور گسترده جهت تولید انرژی الکتریکی توسط سوخت‌های فسیلی ساخته شده‌اند. این موضوع در مرجع [51] تأکید شده است. شکل 6 برخی از این چالش‌ها را نشان می‌دهد [52, 53]. با اعمال شبکه‌های هوشمند، می‌توان برخی از این مشکلات را برطرف نموده و منابع تجدیدپذیر بیشتری را به شبکه متصل کرد. ابزارهای شبکه‌ی هوشمند و فناوری پیاده شده در شبکه‌های برق، عبور جریان دوسویه و مخابره‌ی اطلاعات را میسر می‌سازند. این ویژگی‌ها به کارآمدی، اطمینان پذیری، ارتباط خودکار با دیگر بخش‌ها و امنیت بیشتر منجر می‌گردد [54]. هر چند ویژگی‌هایی همچون، متناوب بودن و برنامه‌ناپذیری این واحدها، همچنان در شبکه وجود خواهند داشت. در این بین، تحقیقات و پژوهش‌های متعددی بر این مطلب متمرکز، و الگوریتم‌ها و راهکارهایی برای برطرف‌سازی این مشکلات ارائه شده است.

## 3- سرمایه‌گذاری و حمایت دولت‌ها

توسعه و تولید انرژی الکتریکی از منابع تجدیدپذیر از جمله راهکارهای مواجهه با مسئله‌ی گرمایش جهانی و کاهش انتشار آلاینده‌های زیست محیطی محسوب می‌شود [55]. همان‌گونه که در بخش‌های قبل اشاره شد، چالش اصلی استفاده از این منابع، هزینه‌ی سرمایه‌گذاری بالای آن‌هاست. بر این اساس لازم است تا منابع مالی مورد نیاز برای سرمایه‌گذاری در این زمینه فراهم گردد. اعطای مشوق‌ها و ایجاد صندوق حمایت مالی از انرژی‌های تجدیدپذیر توسط دولت می‌تواند راهگشا باشد. توسعه انرژی‌های تجدیدپذیر در سال‌های اخیر مورد حمایت بسیاری از کشورها بوده است. برای مثال، در سال 2014 چین با 83,3 میلیارد دلار، بزرگترین سرمایه‌گذاری در بخش انرژی‌های تجدیدپذیر را به خود اختصاص داده است. ایالات متحده با 38,3 میلیارد دلار و ژاپن با 35,7 میلیارد دلار به ترتیب دومین و سومین کشورهای سرمایه‌گذار در این زمینه بوده‌اند [56].

### 3-1- افق برنامه‌ریزی بلند مدت

کشورهای مختلف با تدوین برنامه‌های راهبردی، توسعه‌ی منابع انرژی پاک را در دستور کار خود قرار داده‌اند. ترکیه یک برنامه بنام انرژی تجدیدپذیر ملی را جهت تأمین 30٪

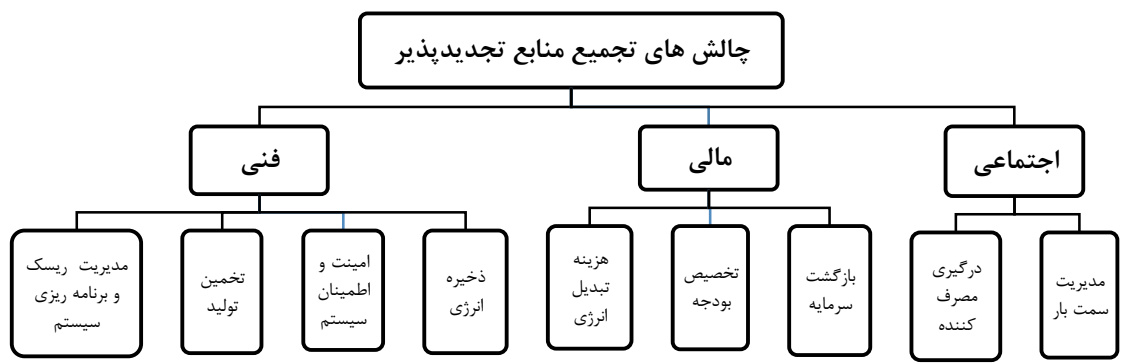

شکل 6: مشکلات استفاده از منابع تجدیدپذیر در شبکه

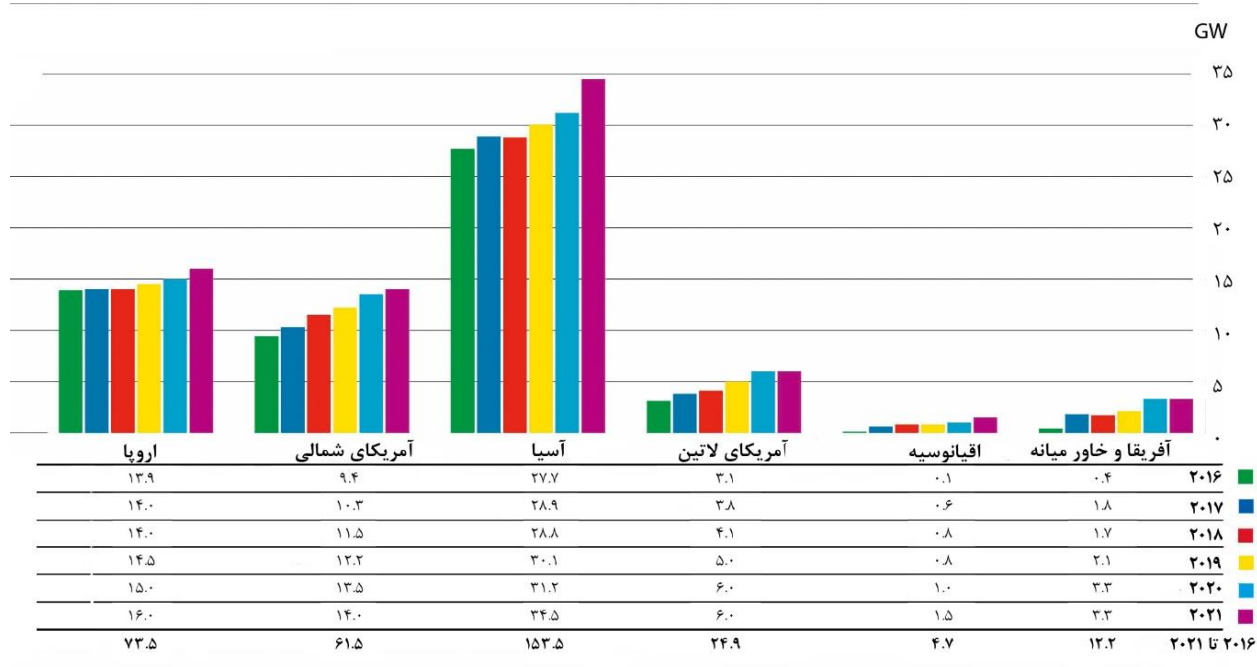

شکل ۷: پیش‌بینی بازار تولید تجدیدپذیرها در مناطق مختلف جهان

تقاضا از طریق این منابع تا سال ۲۰۲۳ ایجاد کرده است. در ماه ژوئن ۲۰۱۵، برزیل توافق‌نامه‌ای را با آمریکا امضا کرد که به موجب آن منابع انرژی تجدیدپذیر این کشور تا سال ۲۰۳۰ به میزان ۲۰٪ افزایش یابند [۵۷]. آلمان مقررات منابع انرژی تجدیدپذیر را در سال ۲۰۰۰ تصویب کرد تا ۵۰٪ از کل انرژی تولیدی در کشور، توسط این منابع تا سال ۲۰۵۰ تأمین گردد [۵۸]. چین برنامه‌ای برای سرمایه‌گذاری ۷۴۰ میلیارد دلار در ۱۰ سال آینده به منظور افزایش سهم منابع پایدار تا سال ۲۰۲۰ تدوین نموده است [۵۹].

در سند چشم‌انداز بیست ساله‌ی ایران نیز، بندهایی به منظور توسعه‌ی این منابع ذکر شده است. فصل دهم ماده‌ی ۶۱ **قانون اصلاح الگوی مصرف** بیان می‌کند: وزارت نیرو موظف است به منظور حمایت از گسترش استفاده از منابع تجدیدپذیر انرژی، شامل انرژی‌های بادی، خورشیدی، زمین گرمایی، آبی کوچک (تا ده مگاوات)، دریایی و زیست توده (مشتمل بر ضایعات کشاورزی، جنگلی، زباله‌ها و فاضلاب شهری، صنعتی، دامی، بیوگاز) و با هدف تسهیل و تجمیع این امور از طریق سازمان‌های ذی‌ربط نسبت به عقد قرارداد بلند مدت خرید تضمینی از تولیدکنندگان غیردولتی برق از منابع تجدیدپذیر اقدام نماید. همچنین ماده ۵۰ **برنامه ششم توسعه** ذکر می‌کند: دولت مکلف است سهم نیروگاه‌های تجدیدپذیر و پاک با اولویت سرمایه‌گذاری بخش غیردولتی (داخلی و خارجی) با حداکثر استفاده از ظرفیت داخلی را تا پایان اجرای قانون برنامه به حداقل ۵٪ ظرفیت برق کشور برساند.

### ۳-۲- بازار تضمین شده

از آنجایی که انرژی‌های تجدیدپذیر به دلیل محدودیت‌های خود نمی‌توانند در بازار انرژی رقابت کنند، فروشندگان باید توسط قانون موظف شوند که بخشی از انرژی خریداری شده را از نیروگاه‌های تجدیدپذیر تامین کنند. نمونه‌ای از این قوانین، تعهد خریداری انرژی غیرفسیلی[1] در انگلستان، قانون انرژی تجدیدپذیر[2] در آلمان و استاندارد جایگاه تجدیدپذیرها[3] در ایالات متحده می‌باشد. قانون NFFO انگلستان و REL آلمان، قیمت مصوبی را برای منابع تجدیدپذیر در بازار تعیین می‌کند تا توان رقابتی با دیگر منابع داشته باشند [۸]. شکل ۷، چشم‌انداز بازار انرژی تجدیدپذیر را در مناطق مختلف جهان نشان می‌دهد.

---

[1] - Non-Fossil Fuel Obligation (NFFO)
[2] - Renewable Energy Law (REL)
[3] - Renewable Protfolio Standard (RPS)

## 4- نتیجه‌گیری و بحث

منابع تجدیدپذیر جایگزینی مناسب برای تأمین انرژی محسوب می‌شوند. پتانسیل عظیم این منابع به قدری است که می‌توانند چندین برابر انرژی مورد نیاز جهان را تأمین نمایند. در این مقاله به معرفی فناوری‌های مرسوم، مزایا و معایب هر یک پرداخته شد. همچنین سهم کشورهای مختلف در استفاده از منابع تجدیدپذیر و تعهد هر کدام در گسترش این منابع عنوان گردید. در کشورهای پیشرو، مازاد هزینه‌ی انرژی‌های نو توسط اعطای یارانه‌های دولتی به منظور نفوذ در بازار جبران می‌شود. این یارانه‌ها می‌تواند از طریق جریمه‌ی نیروگاه‌های سنتی بخاطر نادیده گرفتن اهداف زیست محیطی تأمین شود. در کشورهایی مانند ایران که چرخه‌ی تولید و توزیع انرژی در دست نهادهای دولتی است، حذف تدریجی یارانه انرژی‌های فسیلی، سوق درآمدهای حاصل از آن به تأمین مالی در پروژه‌های تولید و توسعه انرژی‌های تجدیدپذیر، تشویق بخش خصوصی جهت سرمایه‌گذاری و ایجاد تقویت همکاری‌های بین‌المللی جهت توسعه این منابع، می‌تواند راهگشا باشد.

## مراجع